# Descriptive examples of the limitations of Artificial Neural Networks applied to the analysis of independent stochastic data

Henry Navarro[1,] Leonardo Bennun[1]

[1]Applied Physics Laboratory, Department of Physics, Faculty of Physical and Mathematical Sciences, University of Concepción

An Artificial Neural Network (ANN) is a mathematical model inspired by the biological behaviour of neurons and by the structure of the brain, which is used to solve a wide range of problems. A network is reached by connecting several neurons with a specific architecture (Hopfield Networks, Kohonen Networks, Perceptron, and so on), in which neurons learn through a process of self-organization.[1] During the learning process of the ANN, when a data is introduced into the network, just the neuron that has a positive activity inside the proximity will be activated at the exit stage.[1, 2]

There is a wide variety of ANN models, which depend on the objective by which they were created, as well as the practical problem they solve.[3] During these last decades, several inconveniences about ANN applications have been found in the literature,[4,5,6,7] as the performance of the ANN neither have much relation to the amount of acquired information, nor the way the algorithm detects the information,[8] but also with issues like: the selection of the network model, the variables to incorporate on it and the pre-processing of the information that will form the training group.[9]

The ANN depends essentially on the exact information of the system under study, and the methods of training that must be used, as the algorithm of ANN during their training have the ability of identifying unnecessary data.[10] Besides, the routines of training require a huge amount of data to make sure that the results can be statistically precise[11], but a lot of algorithms have been proposed to improve the performance of the ANN.[12, 13] Therefore, it can be seen that it is important to know more about the precision and the sturdiness of the ANN.[14]

Given the wide range of applications of the ANN towards different areas of science (financial and economic modelling, market profiles - customer, medical applications, management of knowledge and discovering of data, optimization of industrial processes and quality control, among others) generate excellent results.[2,15,16] Given the great versatility of the applications and the uses of the ANN, it is possible that can be used in system for which not necessarily has an optimum performance, they may be redundant or inefficient. So, that would be advisable assess the own characteristics of the nature of the problem to solve, in order to assess the application of an ANN.

It can be mentioned that the ANN cannot produce better performance than statistical methods when stochastic independent events are analyzed. As a characteristic example, let´s considering a coin flipped. If a ANN is trained in order to predict the behaviour of the coin, , it will give a correct answer half of the times, and a wrong answer the other half, on average. The ANN cannot perform better than the statistical method, because the system under study is composed of independent stochastic events. In order to analyze results from spectroscopic techniques[7], ANN cannot perform better either, this is because the obtaining of spectroscopic data is ruled by pure stochastic events. This can be described as follows: Let us consider a wide set of absolutely equal photons that impact on the sample. "*A priori*" nobody knows the interaction each one will produce. This interaction is stochastic and it is defined at subatomic levels. The reflectance spectrum of the sample is composed of the answer of the incident photons absolutely equals. This spectrum is composed of a finite set (not infinite) of possibilities. Now, this spectrum will become more complex if heterogeneities of the sample are increased, and there will be more possibilities and different "responses". We can divide this emerging spectrum by the number of incident photons, obtaining then a probability. This probability can be interpreted as the average response of the sample per each incident photon. In other words, the obtaining of spectroscopic data obeys to pure stochastic events. If an ANN is trained in order to improve that possibility, naturally it will not be done, because the spectrum produced is obtained by a sequence of stochastic independent events.

In this letter, it can be also described other systems where the ANN can not either have a better performance when they process independent stochastic events.

Let us suppose that we liberate X air particles in a closed room, and we will also suppose that we want measure the population of particles per unit of volume in the system. Then we take a sample of particles in a small volume, which in our system is assigned with a value of 3 particles. Later we get another sample in which there are 5 particles and again we get one more sample and we observe that there are 9 particles in the same fraction of volume. From the sequence of the collected data, through statistic methods an average value of particles/volume in the system can be obtained. On the other hand, we know that the second law of thermodynamics establishes that entropy (merely a statistic concept) tends to its maximum value, so the system tends to the state of maximum probability, which is consistent with an average population of X/Vol. This result is totally compatible with the fact that the pressure will tend to be constant in its most probable value. If we train an ANN in order to make it evaluate the most probable population of particles/Vol there is in the room, we will note that the efficiency provided by an evaluation purely statistic will not be improved. Another example, based in the second law of thermodynamics, is described as follows:

Let us suppose that we dissolve 5 g of salt (as NaCl) in 10 $cm^3$ of water and the concentration obtained in very small volumes (e.g. at molecular scale) is assessed. Concentration values obtained in every sample, present differences in every measurement. If we intend to train an ANN with microscopic results in order to predict the macroscopic result, such network cannot improve the result of C= 0.5 $g/cm^3$, which is the value obtained from a stochastic composition of the evaluation in small volumes. By means of these two simple examples, we can realize that they clarify the fact that an improvement can not be obtained by the training of an ANN, because the events that took place correspond to stochastic independent events.

In the world of games, specifically in casinos of gambling industry (chance games), we could propose that an ANN can be trained in order to win in the game of roulette/Bank Craps. These games are pure statistical systems. The ANN cannot establish a procedure in order to systematically win in these games, just because the probability of loosing is greater than the probability of winning. If an ANN proposes a particular strategy, in average, this one cannot improve the efficiency in the performance, which is calculated through the corresponding statistic evaluation. Neural networks are approximately 30-40 years old and did not affect (and they will not do it) the profits of the casinos industry.

Now, according to the principle of Maximum Plausibility, there is only one method that produces the best results for the kind of data analysed (spectroscopic or statistic data). The result produced by training an ANN cannot improve the occurrence probability for every number when spectral techniques are measured. In other words, an ANN is not able to produce a method better than the statistic analysis applied to the spectroscopic techniques.

Clearly, an independent stochastic event (spectroscopy) can be described as the best way to use statistically pure methods, and it is not expected that a NN perform better than them, because if there was a way or method that improves statistic methods, it violates the principle of Causality, by predicting results with better efficiency than the statistic methods. Along these descriptive methods, we have been able to indicate that no matter how much we train an artificial neural network ANN in order to improve the probability of pure stochastic events, this will not happen.